\documentclass[reprint, amsmath, amssymb, groupaddress, aps, prb]{revtex4-1}
\usepackage{graphicx}
\usepackage{amsmath}
\begin{document}
\title{Single-Layer Ferromagnetic and Piezoelectric CoAsS with Pentagonal Structure}
\author{Lei Liu}
\author{Houlong L. Zhuang}
\email{zhuanghl@asu.edu}
\affiliation{School for Engineering of Matter Transport and Energy, Arizona State University, Tempe, AZ 85287, USA}
\date{\today}
\begin{abstract}
Single-layer pentagonal materials are an emerging family of two-dimensional (2D) materials that could exhibit novel properties due to the building blocks being pentagons instead of hexagons as in numerous 2D materials. Based on our recently predicted single-layer pentagonal CoS$_2$ that is an antiferromagnetic (AFM) semiconductor, we replace two S atoms by As atoms in a unit cell to form single-layer pentagonal CoAsS. The resulting single-layer material is dynamically stable according to the phonon calculations. We find two drastic changes in the properties of single-layer pentagonal CoAsS in comparison with those of CoS$_2$. First, we find a magnetic transition from the AFM to FM ordering. We understand that the transition is caused by the lower electronegativity of As atoms, leading to the weakened bridging roles on the superexchange interactions between Co ions. Single-layer pentagonal CoAsS also shows significantly stronger magnetocrystalline anisotropy energy due to stronger spin-orbit coupling. We additionally perform Monte Carlo simulations to calculate the Curie temperature of single-layer pentagonal CoAsS and the predicted Curie temperature is 103 K. Second, we find that single-layer pentagonal CoAsS exhibits piezoelectricity, which is absent in single-layer pentagonal CoS$_2$ due to its center of symmetry. The computed piezoelectric coefficients are also sizable. The rare coexistence of FM ordering and piezoelectricity makes single-layer pentagonal CoAsS a promising multifunctional 2D material.
\end{abstract}
\maketitle
\section{Introduction}
Multifunctional materials have been immensely studied due to their combinations of two or more chemical and physical properties that can be used and tailored for specific applications.\cite{sasabe2010multifunctional,cobley2011gold,cha2013carbon} For example, various biomedical applications of gold nanostructures have been realized ranging from diagnostics to medical imaging mainly due to the functional properties such as localized surface plasmon resonance.\cite{cobley2011gold} Multifunctional properties are also desirable for 2D materials. For instance, 2D oxide thin films that possess magnetism, ferroelectricity, and thermoelectricity simultaneously, have shown great potential in electronic device applications.\cite{dawber2005physics, bowen2007using, ohta2007giant}

Among a number of functional properties, magnetism is certainly an important one that makes a 2D material attractive and promising for spintronic applications. A number of 2D materials have been predicted to exhibit ferromagnetic (FM) or antiferromagnetic (AFM) orderings.\cite{cheng2013prediction,li2014crxte,cao2015tunable} Some of these predictions such as single-layer CrI$_3$ and Fe$_3$GeTe$_2$ have also been confirmed in recent experiments.\cite{huang2017layer, fei2018two, deng2018gate}

As another critical functional property for a semiconducting 2D material, piezoelectricity describes a physical phenomenon that couples mechanical strains and electric fields. Reed {\it et al.} first performed theoretical studies of piezoelectricity on several 2D transition-metal dichalcogenides.\cite{piezo,piezo2} A recent experiment on single-layer MoS$_2$ has confirmed the theoretical predictions.\cite{wu2014piezoelectricity}

But few of the above examples of 2D materials have shown the coexistence of magnetism and piezoelectricity. The former property often involves a transition-metal element being a component of a 2D material and the latter requires the absence of a center of symmetry. Searching for multifunctional 2D materials with both magnetism and piezoelectricity is therefore critical to expand the functional applications of 2D materials.

\begin{figure}
  \includegraphics[width=8cm]{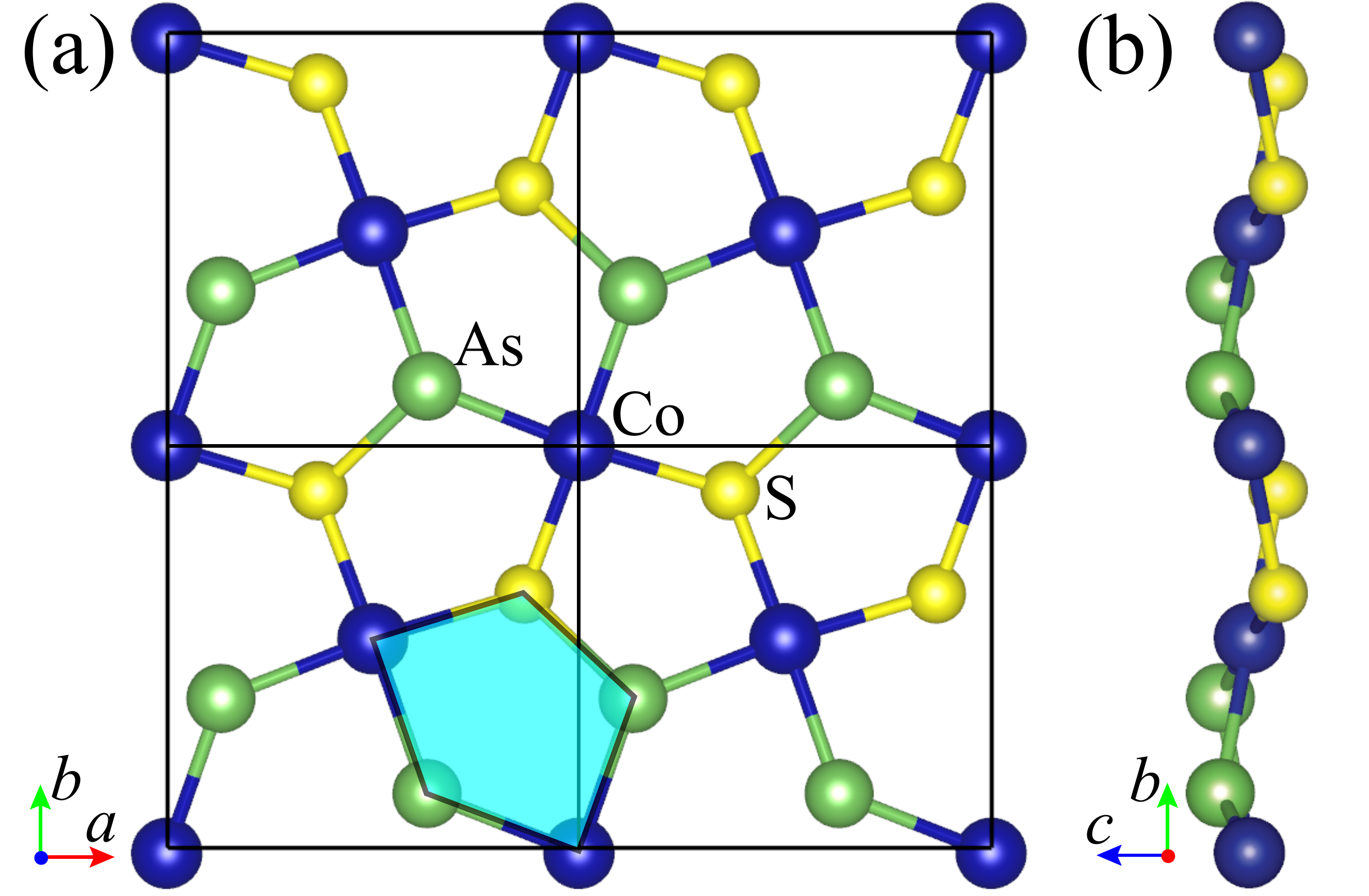}
  \caption{(a) Top and (b) side views of a 2 $\times$ 2 $\times$ 1 supercell of single-layer CoAsS with pentagonal structure. A type 2 pentagon encloses the cyan shaded area shown in (a).}
  \label{fig:structure}
\end{figure}
To discover novel 2D materials with both magnetism and piezoelectricity, we adopt the strategy of exploring  structure-properties relationships by modifying the building blocks of a 2D material from hexagons to pentagons. In particular, we used pentagonal geometries in conjunction with density functional theory (DFT) calculations to predict novel 2D materials.\cite{liu2018encoding,liu2018can} Furthermore, through datamining the Materials Project,\cite{Jain2013} we identified a series of pyrites structures with embedded type 2 pentagons,\cite{liu2018single} one of which is illustrated in Fig.\ref{fig:structure}(a). Tessellating this type of pentagons in a plane forms the Cairo tessellation, a ubiquitous pattern seen on the Cairo street.\cite{wells1991penguin} We recently studied the electric structure and magnetic properties of single-layer CoS$_2$ with the Cairo tessellation. Although this single-layer pentagonal material exhibits the AFM ordering, the structure remains to have an inversion center, prohibiting the occurrence of the piezoelectric effect. We therefore aim to design a 2D multifunctional material based on single-layer pentagonal CoS$_2$ by modifying the structure to achieve the combination of magnetism and piezoelectricity.

To break the inversion symmetry of single-layer CoS$_2$, we partially replacing the S atoms bridging the Co atoms by As atoms to form single-layer CoAsS. We expect the substitution will introduce the piezoelectricity. Figure~\ref{fig:structure} shows the top and side views of a 2~$\times$~2~$\times$~1 supercell of single-layer CoAsS. Each unit cell consists of the same number (2) of Co, As, and S atoms.  Similar to single-layer pentagonal CoS$_2$, single-layer pentagonal CoAsS has a bulk counterpart, which is a type of sulfarsenide minerals that include NiAsS and FeAsS.\cite{saintilan2017re} Bulk CoAsS is also an example of arsenopyrite-like structure with a general chemical formula $AB$S ($A$ = Fe, Co, Ni; $B$ = As, Sb).\cite{bachhuber2014phase} Although bulk CoAsS was successfully synthesized in the 1970s\cite{nahigian1974preparation} and recently studied using DFT calculations,\cite{bachhuber2014phase} single-layer CoAsS has not yet been obtained or computationally characterized. A DFT + $U$ study of single-layer CoAsS therefore will stimulate interest in this new 2D material.

%----------------------------------------------------------------------
\section{Methods}
 We use the Vienna {\it Ab-initio} Simulation Package (VASP, version 5.4.4)\cite{Kresse96p11169} to perform all the DFT calculations. We also use the Perdew-Burke-Ernzerhof (PBE) functional for approximating the exchange-correlation interactions.\cite{Perdew96p3865} Furthermore, we use the standard PBE version of the potential datasets for Co, As, and S generated based on the projector-augmented wave (PAW) method.\cite{Bloechl94p17953,Kresse99p1758} The potential datasets treat the 3$d^8$ and 4$s^1$ electrons of Co atoms, the 4$s^2$ and 4$p^3$ electrons of As atoms, and the 3$s^2$ and 3$p^4$ electrons of S atoms as valence electrons. The plane waves with the cut-off kinetic energy of 550 eV are used to approximate the electron wave functions. We adopt a $\Gamma$-centered $12~\times~12~\times~1$ Monkhorst-Pack $k$-point grid to sample the $k$ points in the reciprocal space.\cite{PhysRevB.13.5188} We keep using the same effective $U$ parameter ($U_\mathrm{eff}$ = 3.32 eV) with the Dudarev method\cite{PhysRevB.57.1505} to treat the $d$ orbitals of Co atoms. This parameter has been shown to lead to similar energy differences between the AFM and FM structures of single-layer pentagonal CoS$_2$ using the PBE+$U$ and HSE06 methods.\cite{liu2018single} The vacuum spacing of the surface slab is set to 18.0~\AA~to avoid image interactions between neighbouring nanosheets of CoAsS. During the VASP calculations, the in-plane lattice constants along with atomic coordinates are fully optimized until the residual forces reach a threshold value of 0.01 eV/\AA.
%------------------------
\section{Results and Discussion}
\begin{table}[b]
  \caption{Energies $\Delta E$ (in meV per formula unit) and magnetic moments $m$ (in $\mu_\mathrm{B}$ per Co ion), in-plane lattice constants $a$ and $b$ (in \AA), and band gaps $E_\mathrm{g}$ (in eV) of single-layer CoAsS with different spin states. The energies are calculated using the energy of the LS-FM state as the reference. HS, IS, LS, AFM, FM, and NM stand for high spin, intermediate spin, low spin, antiferromagnetic, ferromagnetic, and non-magnetic, respectively. All these results are obtained from the PBE + $U$ ($U_\mathrm{eff}$ = 3.32 eV) calculations.}
  \begin{ruledtabular}
    \begin{center}
      \begin{tabular}{cccccc}
            Spin state & $\Delta E$& $m$ &$a$  &$b$  & $E_\mathrm{g}$  \\
            \hline
HS-AFM&45&0.00&5.69&5.52&0.32\\
HS-FM&0&2.00&5.75&5.69&0.34\footnote{Spin-up component},0.64\footnote{Spin-down component}\\
IS-AFM&45&0.00&5.69&5.52&0.32\\
IS-FM&0&2.00&5.75&5.69&0.32$^\mathrm{a}$,0.64$^\mathrm{b}$\\
LS-AFM&149&0.00&5.65&5.61&0.09\\
LS-FM&0&2.00&5.75&5.69&0.34$^\mathrm{a}$,0.64$^\mathrm{b}$\\
NM&228&0.00&5.64&5.64&0.48\\ 
      \end{tabular}
    \end{center}
  \end{ruledtabular}
  \label{summary}
\end{table}

We follow the same procedure applied to single-layer pentagonal CoS$_2$ to determine the ground-state magnetic ordering of single-layer pentagonal CoAsS.\cite{liu2018single} Namely, we assume three different initial magnetic moments (5, 3, and 1 $\mu_\mathrm{B}$, respectively) for Co ions to form high-spin (HS), intermediate-spin (IS), and low-spin (LS) states, for each of which we account for the AFM and FM arrangements of the two Co ions in a unit cell. We also perform non-spin polarized DFT calculations and the resulting state is denoted as non-magnetic (NM). Table~\ref{summary} summarizes the energies of single-layer pentagonal CoAsS adopting the seven different states. As can be seen, the NM state exhibits the highest energy. More importantly, all the FM states are relaxed into the same energy and in-plane lattice constants, independent of the initially speculated spin states. By contrast, there are two different AFM states. The HS-AFM and IS-AFM states have the same energy, whereas the LS-AFM state exhibits a much higher energy. But both energies are higher than the energies of FM states. We therefore conclude that the FM state is the ground state of single-layer pentagonal CoAsS, and HS/IS-AFM state is the second lowest-energy state. We henceforth focus on single-layer pentagonal CoAsS with the FM ordering. 

We also see from Table~\ref{summary} that each Co ion has an integer magnetic moment of 2 $\mu_\mathrm{B}$, showing localized magnetism in contrast to itinerant magnetism exhibited by {\it e.g.,} single-layer Fe$_3$GeTe$_2$.\cite{PhysRevB.93.134407} Table~\ref{summary} also lists the in-plane lattice constants of single-layer pentagonal CoAsS. As expected, the in-plane lattice constants increase to 5.75 \AA~ and 5.69 \AA~from 5.34 \AA~ and 5.43 \AA, respectively, as in single-layer pentagonal CoS$_2$\cite{liu2018ptp} due to the larger radius of As atoms.

\begin{figure}
  \includegraphics[width=8cm]{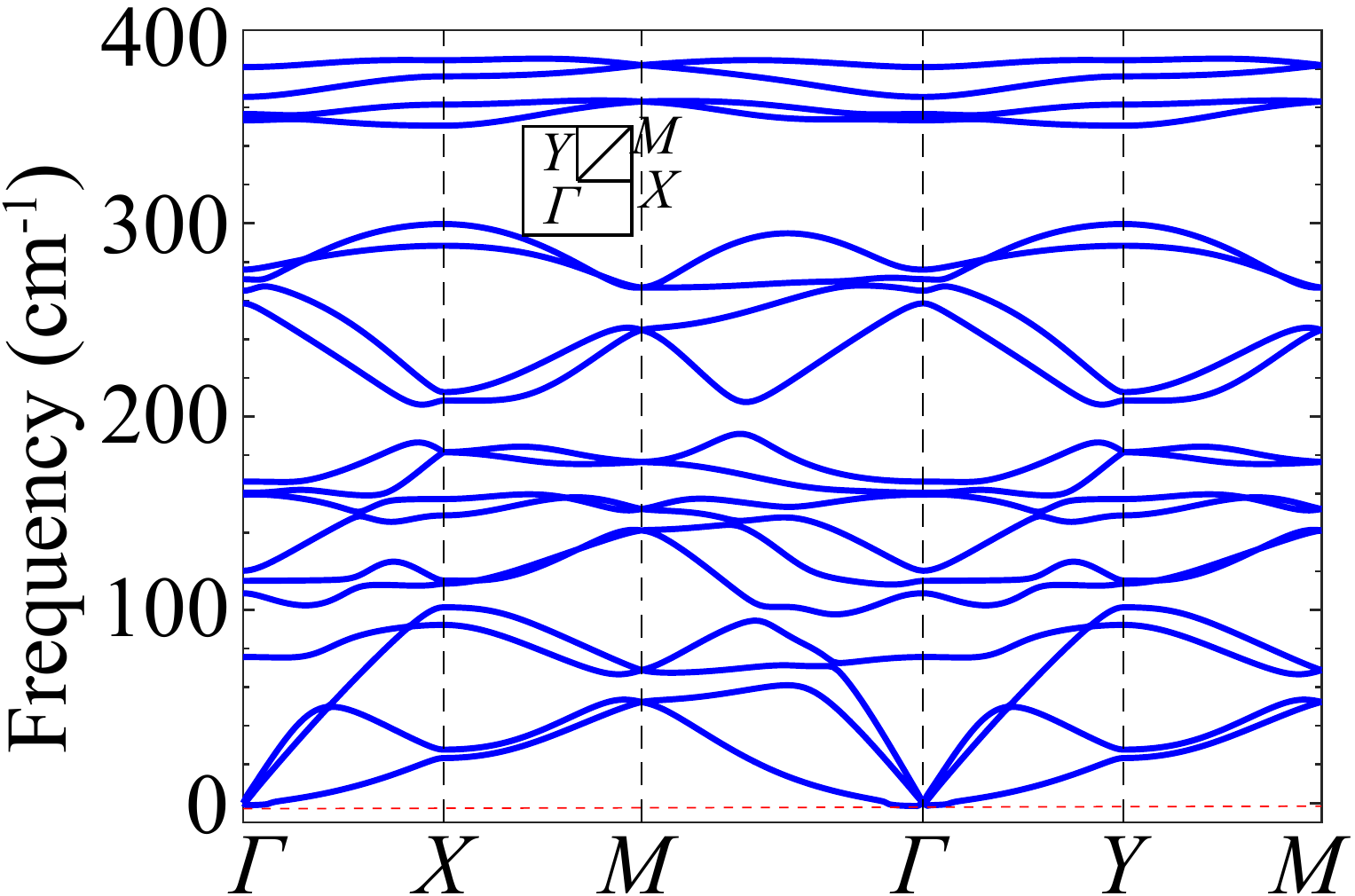}
  \caption{Computed phonon spectrum of single-layer pentagonal CoAsS using 4 $\times$ 4 $\times$ 1 supercells.}
  \label{fig:phonon}
\end{figure}

To examine the effects on the dynamical stability caused by partially replacing S by As atoms, we calculate the phonon spectrum of single-layer pentagonal CoAsS. Figure~\ref{fig:phonon} shows the computed phonon spectrum using the force constants of 4 $\times$ 4 $\times$ 1 supercells. We observe that, similar to single-layer pentagonal CoS$_2$\cite{liu2018ptp}, single-layer pentagonal CoAsS is dynamically stable ({\it i.e.,} no imaginary frequencies) except that the maximum phonon frequencies at different $k$ points is reduced because of the heavier As atoms. For example, at the {\it $\Gamma$} point the maximum vibrational frequency is 381 cm$^{-1}$ for CoAsS, in contrast to 455 cm$^{-1}$ for CoS$_2$ at the same $k$ point.

\begin{figure}
  \includegraphics[width=8cm]{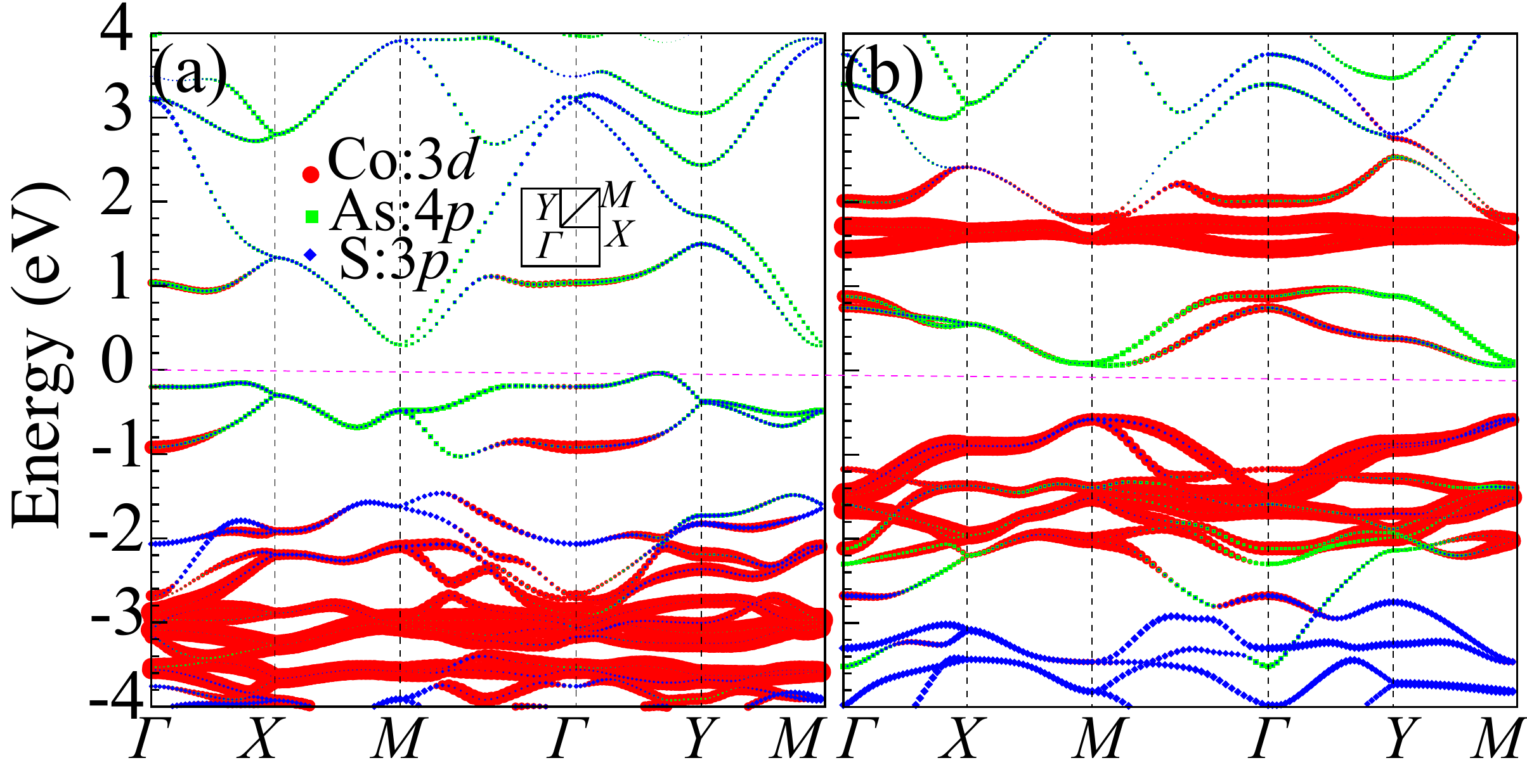}
  \caption{(a) Spin-up and (b) spin-down orbital-resolved band structures of single-layer pentagonal CoAsS calculated with the PBE + $U$ ($U_\mathrm{eff}$ = 3.32 eV) method. The inset of (a) shows the high- symmetry $k$-point path.}
  \label{fig:bandstructure}
\end{figure}

Figure~\ref{fig:bandstructure} shows the orbital-resolved band structures for the spin-up and spin-down electrons, revealing that single-layer pentagonal CoAsS is a semiconductor with the spin-up and spin-down bandgaps of 0.32 and 0.64 eV, respectively. The spin-up bandgap is an indirect bandgap with the conduction band minimum (CBM) at the $M$ point and the valence band minimum (VBN) between the $\Gamma$ and $Y$ points, whereas the spin-down bandgap is a direct one with both the CBM and VBM at the $M$ point. Three types of orbitals (3$d$ orbitals of Co; 4$p$ and 3$p$ orbitals of As and S, respectively.) dominate the bands near the bandgaps. In particular, the CBMs at the $M$ points of the spin-up and spin-down bandgaps are dominated from the 4$p$ orbitals of As. For the spin-up band structure, 4$p$ orbitals of As and 3$p$ orbitals of S are mixed to form the VBM. For the spin-down band structure, more visible contributions to the VBM result from the mixed 3$d$ orbitals of Co and 3$p$ orbitals of S. These mixed orbitals contribute to the exchange interactions between Co ions leading to the FM ordering.

We also apply the HSE06 hybrid density functional to calculate the bandgaps of the optimized structure from the PBE + $U$ ($U_\mathrm{eff}$ = 3.32 eV) method. We find that the HSE06 functional results in the same conclusion that single-layer pentagonal CoAsS is a semiconductor and the FM structure is more stable than the AFM structure by 35.85 meV per formula unit. This energy difference is also similar to that (44.91 meV per formula unit) using the PBE + $U$ ($U_\mathrm{eff}$ = 3.32 eV) method. Furthermore, the spin-up and spin-down bandgaps using the HSE06 functional are 0.29 and 1.64 eV, respectively.

To understand the mechanism of the transition from single-layer pentagonal, AFM CoS$_2$ to FM CoAsS, we note that the Co-S-Co superexchange interactions in single-layer pentagonal, AFM CoS$_2$ are rather weak and the calculated exchange integral is 3.01 meV leading to the low N$\acute{e}$el temperature. Unlike the Co-Co exchange interactions in single-layer pentagonal CoS$_2$ bridged only by the S ions, the Co-Co exchange interactions in single-layer pentagonal CoAsS are bridged by both As and S ions. This is reflected by the orbital mixing in the orbital-resolved band structure shown in Fig.\ref{fig:bandstructure}.
\begin{figure}
  \includegraphics[width=8cm]{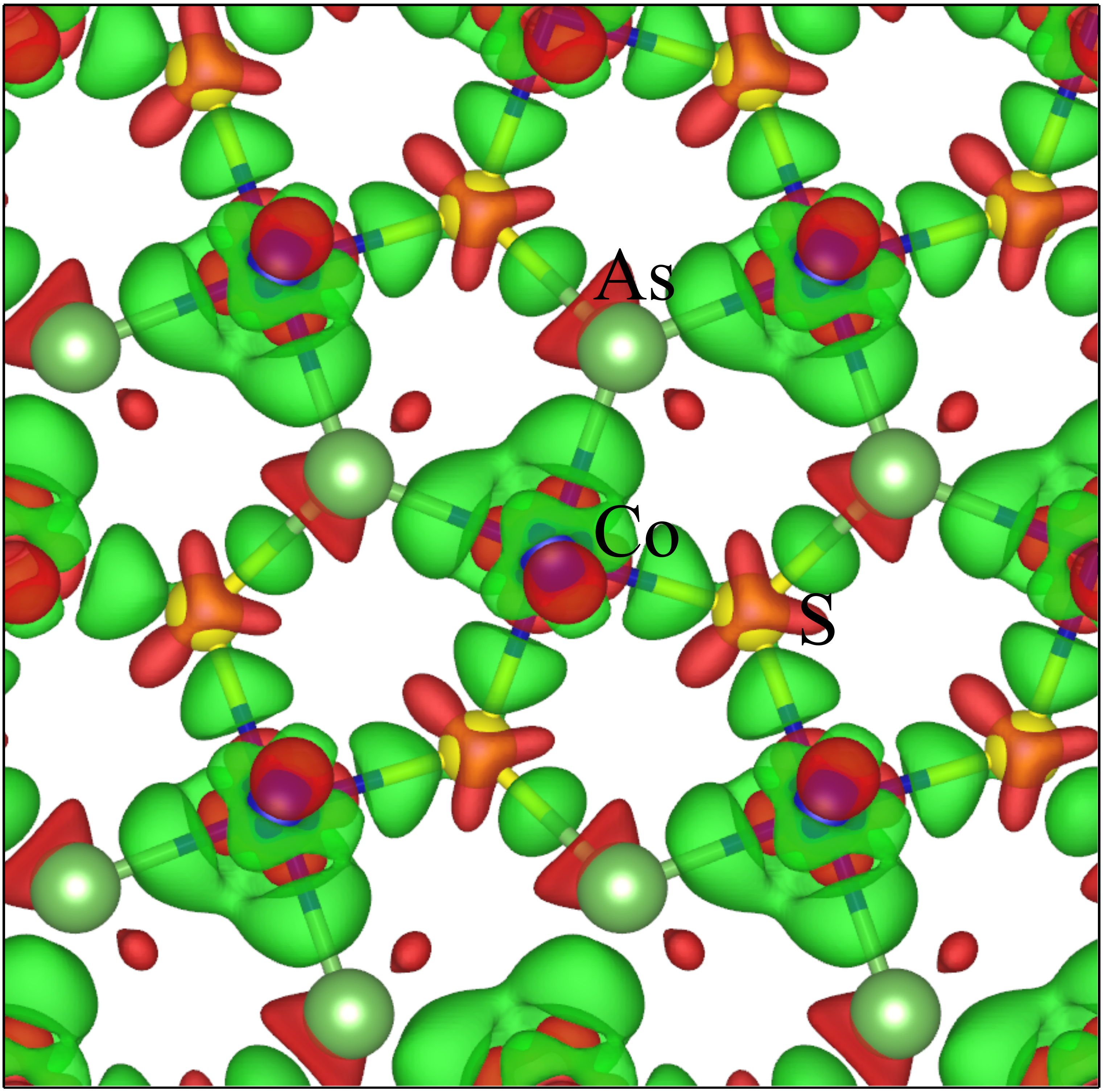}
  \caption{Charge density difference of single-layer pentagonal CoAsS. Green and red isosurfaces refer to charge depletion and accumulation, respectively. The isosurface value is 0.007 $e$/$a_0^3$ ($a_0$: Bohr radius).}
  \label{fig:charg}
\end{figure}

The electrons surrounding the As and S ions in single-layer pentagonal CoAsS play the roles of bridging the exchange interactions between the Co ions. The magnitude of electron density around an As or S ion therefore qualitatively determines the ``bridge width" describing the strength of exchange interactions. Intuitively, the larger ``bridge width" corresponds to stronger exchange interactions. To compare the ``bridge width" of As and S ions, Fig.\ref{fig:charg} displays a plot of the charge density difference $\Delta \rho$, which is calculated as the difference of the charge density of single-layer pentagonal CoAsS with reference to the total charge density of isolated Co, As, and S atoms. The $\Delta \rho$ plot shows that electron accumulates more significantly near the S ions than near the As ions, consistent with the higher electronegativity of S (2.5) over As (2.0) on the Pauling scale.\cite{pauling1960nature} The narrower ``bridge width" results in the weakened Co-As-Co superexchange interactions that would otherwise enhance the AFM ordering. Consequently, the spins in Co ions prefer an alignment in the FM manner, competing with the AFM alignment owing to the Co-S-Co superexchange interactions. 

To further demonstrate the importance of electronegativity in affecting the magnetic ordering of single-layer pentagonal CoAsS, we perform an energy calculation on an imaginary 2D material CoOS. Namely, the As atoms are replaced by more electronegative O atoms (3.5 on the Pauling scale).\cite{pauling1960nature} As a result of this replacement, the energy of CoOS with the AFM ordering is lower than that with the FM ordering by 86.44 meV per formula unit. As a reference, for single-layer pentagonal CoS$_2$, the corresponding energy difference is merely 11.99 meV per formula unit.\cite{liu2018ptp} We therefore conclude that the electronegativity of the atoms bridging the Co ions plays a critical role in affecting the magnetic ordering. As such, one may use the electronegativity of bridging ions as a metric to assess the exchange coupling strength in 2D magnets.

\begin{figure}
  \includegraphics[width=8cm]{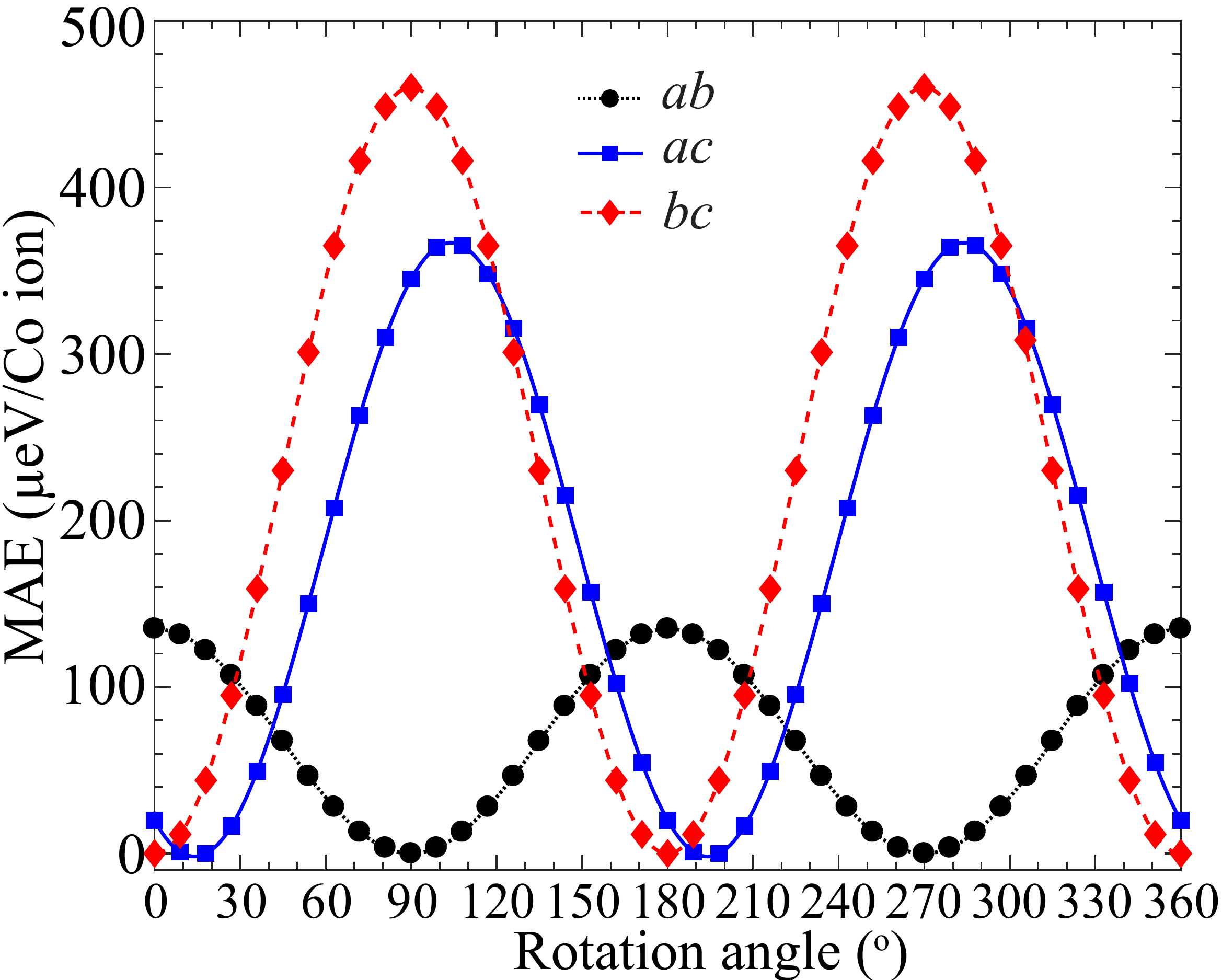}
  \caption{Magnetocrystalline anisotropy energy of single-layer pentagonal CoAsS in the $ab$, $ac$, and $bc$ planes. The energy minimum is set to zero for each plane.}
  \label{fig:mae}
\end{figure}

Having understood the mechanism of magnetic transition, we set to evaluate the magnetocrystalline anisotropy energy (MAE) of single-layer pentagonal CoAsS. To calculate the MAE, we consider the spin-orbit coupling (SOC) in the calculations.\cite{PhysRevB.93.224425} We set the parallel spin vectors of the two Co ions initially along the $a$ or $b$ axis and gradually change the angle of these two vectors with reference to one of these two axes in the $ab$ and $ac$, and $bc$ planes, respectively. The MAE is therefore defined as the energy at different rotation angles subtracting the minimum energy in each of the three planes. Figure~\ref{fig:mae} shows the calculated MAE as a function of rotation angle in the $ab$, $ac$, and $bc$ planes. In the $ab$ plane, the minimum-energy spin orientation is along the $b$ axis, and the highest MAE (135 $\mu$eV/Co ion) occurs when the spin vectors are along the $a$ axis. In the $ac$ plane, the minimum-energy spin orientation is along the direction that has an angle of 10$^\circ$ about the $a$ axis. The corresponding largest MAE is (365 $\mu$eV/Co ion). In the $bc$ plane, the MAEs show the largest energy variation, from the energy minimum along the $b$ axis to the energy maximum (460 $\mu$eV/Co ion) along the $c$ axis. This maximum MAE is much higher than that (153 $\mu$eV/Co ion) of single-layer pentagonal CoS$_2$ due to the presence of heavier As ions associated with stronger SOC. Strong magnetocrystalline anisotropy is a necessary condition for single-layer pentagonal CoAsS to exhibit long-range magnetic ordering.\cite{PhysRevLett.17.1133, paul2017computational} 

\begin{figure}
  \includegraphics[width=8cm]{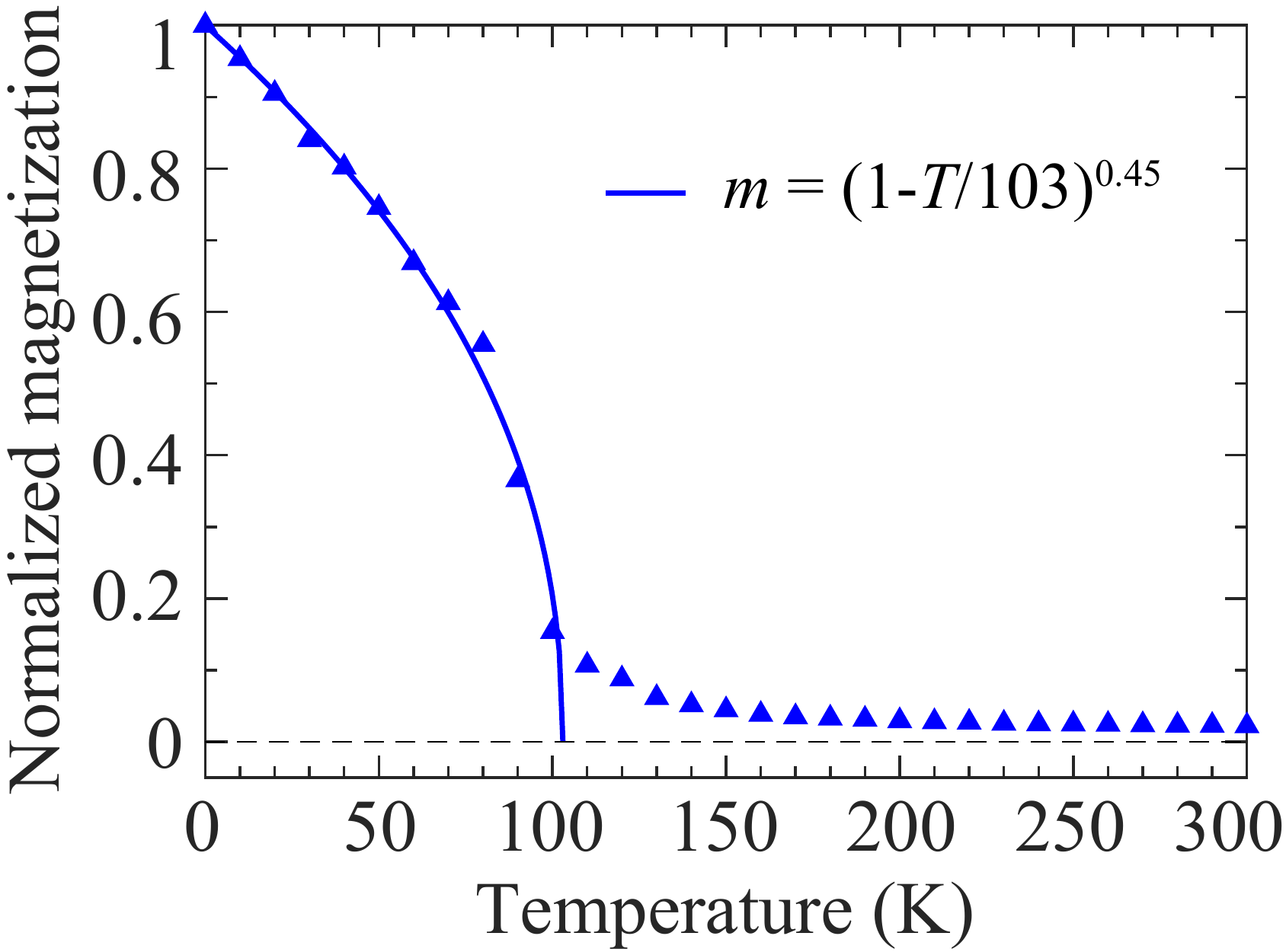}
  \caption{Normalized magnetization $m$ (represented by blue triangles) as a function of temperature $T$ in single-layer ferromagnetic, pentagonal CoAsS obtained from Monte Carlo simulations. The magnetization is fitted to the equation: $m(T) = \Big(1-\frac{T}{T_\mathrm{C}}\Big)^\beta$, plotted as a solid blue line.}
  \label{fig:mc}
\end{figure}

\begin{table*}[!htbp]
  \caption{In-plane elastic constants $C_{11}$ (N/m), $C_{22}$ (N/m), and $C_{12}$ (N/m), and piezoelectric coefficients $e_{11}$ (10$^{-10}$ C/m), $e_{12}$ (10$^{-10}$ C/m), $d_{11}$ (pm/V), $d_{12}$ (pm/V), and $d_{26}$ (pm/V).}
  \begin{ruledtabular}
    \begin{center}
      \begin{tabular}{cccccccccc}
             $C_{11}$ &$C_{12}$ &$C_{66}$ &$C_{22}$ & $e_{11}$ & $e_{12}$  & $e_{26}$ & $d_{11}$ & $d_{12}$ & $d_{26}$ \\
            \hline
    75.44&6.17&26.94&51.53 & -2.47 & 1.38 &1.09 & -3.52 & 3.10 & 2.02\\
      \end{tabular}
    \end{center}
  \end{ruledtabular}
  \label{piezo}
\end{table*}

We next employ the VAMPIRE package\cite{evans2014atomistic} to compute the Curie temperature of single-layer pentagonal CoAsS via Monte Carlo (MC) simulations. The simulations are based on the Heisenberg model with the exchange Hamiltonian for the exchange interaction of a system\cite{evans2014atomistic}
\begin{equation}
\textrm{H} = -\frac{1}{2}\sum_{i \neq j}J_{ij} S_i  S_j,
\end{equation}
where $J_{ij}$ is the exchange integral between adjacent spins of Co ions, and $S_i$ and $S_j$ are the spin moment ({\it i.e.} 1.0 $\mu_\mathrm{B}$) of the $i$-th and $j$-th Co ions, respectively. Following the convention, positive exchange integral $J_{ij}$ denotes FM ordering. We consider only the nearest-neighbour interactions between Co ions and calculate the exchange integral from the energy difference between the AFM and FM configurations as
\begin{equation}
J = \frac{E_{\mathrm{AFM}}-F_{\mathrm{FM}}}{8},
\end{equation}
where $E_{\textrm{AFM}}$ and $E_{\textrm{FM}}$ are the energies of a unit cell of single-layer pentagonal CoAsS with the same structure and different magnetic configurations (AFM and FM, respectively). We obtain a larger $J$ (12.77 meV) than that (3.01 meV) of single-layer pentagonal CoS$_2$.\cite{liu2018single} 

To perform the MC calculations, we construct a supercell of single-layer CoAsS with the in-plane size of 25 nm $\times$ 25 nm, sufficiently large to minimize the statistical noise caused by the finite-size effects. We apply the periodic boundary conditions in both the $a$ and $b$ directions. Figure~\ref{fig:mc} displays temperature-dependent normalized magnetization using $10^4$ equilibration and $10^4$ averaging steps. We fit the results to the equation\cite{evans2014atomistic}
\begin{equation}
m(T) = \Big(1-\frac{T}{T_\mathrm{C}}\Big)^\beta,
\end{equation}
and we obtain an estimated Curie temperature $T_\mathrm{C}$ of 103K and the critical exponent $\beta$ of 0.45. Although the $T_\mathrm{C}$ is still below room temperature, it appears higher than the N$\acute{e}$el temperature ($\sim 20$K \cite{liu2018single}) of single-layer CoS$_2$ due to the enhanced exchange interactions.

In addition to the magnetic transition from single-layer AFM CoS$_2$ to FM CoAsS, another important difference between these two single-layer materials is their symmetries. We use the Phonopy package\cite{phonopy} to perform a symmetry analysis and to identify the space groups of the surface slabs of single-layer pentagonal CoS$_2$ and CoAsS. The space group numbers of these two single-layer materials are 14($P2_1/c$) and 7($Pc$), respectively. A symmetry change is likely to cause novel properties such as the piezoelectricity to occur in 2D materials like MoS$_2$.\cite{doi:10.1021/jz3012436} Due to the presence of the inversion symmetry, single-layer pentagonal CoS$_2$ shows no piezoelectric effect. We also confirm this in our calculations. By contrast, the inversion center is absent in single-layer pentagonal CoAsS, as can be seen in Fig.\ref{fig:structure}(a). 

The strength of piezoelectricity is quantified by piezoelectric coefficients. To calculate the piezoelectric coefficients of single-layer pentagonal CoAsS, we first compute the elastic constants with a symmetry-general method.\cite{PhysRevB.65.104104} The surface slab is a monoclinic structure. It is therefore associated with 13 independent elastic constants.\cite{nye1985physical} For a 2D material, we only need to focus on the elastic constants (a four-rank tensor) that contains no index of 3 (the $z$ component). As a result, there are four independent elastic constants of interest: $C_{1111}$, $C_{1122}$, $C_{1212}$, and $C_{2222}$, which are written as $C_{11}$, $C_{12}$, $C_{66}$, and $C_{22}$ in the Voigt notation.\cite{voigt2014lehrbuch} Table~\ref{piezo} shows the four computed elastic constants. We convert the dimension from N/m$^2$ to a common dimension N/m for 2D materials by multiplying the obtained elastic constants from VASP calculations with the $z$ length (18.0~\AA) of the surface slab. By doing this, we remove the dependence of the computed elastic constants on the vacuum spacing. The same unit transformation is also applied to the piezoelectric coefficients (see below). 

We next calculate the piezoelectric coefficients $e_{ijk}$ and $d_{imn}$ of single-layer pentagonal CoAsS. These two sets of piezoelectric coefficients are related via the following equation using the Einstein notation:\cite{doi:10.1021/jz3012436, doi:10.1021/acsnano.5b03394, doi:10.1063/1.4934750}
\begin{equation}
e_{ijk} = d_{imn}C_{mnjk}.
\end{equation}
Due again to the 2D nature, only the components with indices 1 and/or 2 are of interest. Moreover, because of the symmetry of the surface slab, only three coefficients $e_{111}$, $e_{122}$, and $e_{212}$ are independent. These three coefficients are calculated as
\begin{equation}
e_{111} = d_{111}C_{1111}+d_{122}C_{2211}+d_{112}C_{1211}+d_{121}C_{2111},
\label{eq1}
\end{equation}

\begin{equation}
e_{122} = d_{111}C_{1122}+d_{122}C_{2222}+d_{112}C_{1222}+d_{121}C_{2122},
\label{eq2}
\end{equation}
and
\begin{equation}
e_{212} = d_{211}C_{1112}+d_{222}C_{2212}+d_{212}C_{1212}+d_{221}C_{2112},
\label{eq3}
\end{equation}
respectively. Reducing Eqs.~\ref{eq1}, \ref{eq2}, and~\ref{eq3} with two-index notations gives rise to $d_{11}$, $d_{12}$, and $d_{26}$ in the following equations that are previously used for other 2D materials:\cite{doi:10.1021/jz3012436, doi:10.1021/acsnano.5b03394, doi:10.1063/1.4934750}
\begin{equation}
d_{11} = \frac{e_{11}C_{22}-e_{12}C_{12}}{C_{11}C_{22}-C_{12}^2},
\label{eq4}
\end{equation}

\begin{equation}
d_{12} = \frac{e_{12}C_{11}-e_{11}C_{12}}{C_{11}C_{22}-C_{12}^2},
\end{equation}
and

\begin{equation}
d_{26} = \frac{e_{26}}{2C_{66}}.
\end{equation}
To obtain $e_{ijk}$, we calculate the polarizations from both electronic and ionic contributions using density functional perturbation theory.\cite{gajdovs2006linear} Table~\ref{piezo} shows that the calculated piezoelectric coefficients are sizable. For example, the computed $d_{11}$ (-3.52 pm/V) is significantly larger than that of a variety of single-layer materials such as MoS$_2$ (1.50 pm/V) and WS$_2$ (1.93 pm/V) with the 2$H$ structure.\cite{duerloo2012intrinsic} The coefficients $d_{11}$ and $e_{11}$ of single-layer CoAsS also exceed all the engineered piezoelectric coefficients of atom doped graphene.\cite{piezo2} Such a significant piezoelectric effect endows single-layer pentagonal CoAsS with an additional functional property besides the existing FM ordering. It might lead to potential applications in sensors, energy conversion and electronics.\cite{kingon2005lead,wu2014piezoelectricity}
%----------------------------------------------------------------------
\section{Conclusions}
 In summary, we have predicted a single-layer ternary compound CoAsS with pentagonal structure using DFT + $U$ calculations. In comparison with single-layer pentagonal CoS$_2$ with the AFM ordering, single-layer pentagonal CoAsS exhibits the FM ordering and significantly stronger MAEs. We suggest that electronegativity plays an important role in leading to the magnetic transition from the AFM to the FM ordering. In addition to the FM ordering, we find single-layer pentagonal CoS$_2$ possesses piezoelectricity with sizable piezoelectric coefficients. Our prediction shows that this novel single-layer pentagonal material may be useful for a variety of applications owing to its multifunctional properties.
%----------------------------------------------------------------------
\begin{acknowledgments}
We thank the start-up funds from Arizona State University (ASU). This research used computational resources of the Agave Research Computer Cluster of ASU and the Texas Advanced Computing Center under Contracts No.TG-DMR170070. 
\end{acknowledgments}
\bibliography{references}
\end{document}